\documentclass[conference]{IEEEtran}
\usepackage{cite}
\usepackage{graphicx}
\usepackage{wrapfig}
\usepackage{color}
\usepackage{float}
\usepackage{latexsym}
\usepackage{caption}
\usepackage{subcaption}
\usepackage{setspace}

\ifCLASSINFOpdf
\else
\fi

\usepackage{stfloats}

\newcommand{\crowdsensing}{\textit{crowdsensing }}
\newcommand{\osensing}{\textit{opportunistic sensing }}

\begin{document}
%
\title{Efficient Opportunistic Sensing using Mobile Collaborative Platform MOSDEN}

\author{\IEEEauthorblockN{Prem Prakash Jayaraman\IEEEauthorrefmark{1},
Charith Perera,
Dimitrios Georgakopoulos and
Arkady Zaslavsky}
\IEEEauthorblockA{CSIRO Computational Informatics\\
Canberra, Australia 2601\\ 
Email: \{prem.jayaraman, charith.perera, dimitrios.georgakopoulos, \\arkady.zaslavsky\}@csiro.au}
\IEEEauthorrefmark{1}Corresponding Author}


\maketitle

\begin{abstract}
Mobile devices are rapidly becoming the primary computing device in people's lives. Application delivery platforms like Google Play, Apple App Store have transformed mobile phones into intelligent computing devices by the means of applications that can be downloaded and installed instantly. Many of these applications take advantage of the plethora of sensors installed on the mobile device to deliver enhanced user experience. The sensors on the smartphone provide the opportunity to develop innovative mobile \osensing applications in many sectors including healthcare, environmental monitoring and transportation. In this paper, we present a collaborative mobile sensing framework namely Mobile Sensor Data EngiNe (MOSDEN) that can operate on smartphones capturing and sharing sensed data between multiple distributed applications and users. MOSDEN follows a component-based design philosophy promoting reuse for easy and quick \osensing application deployments. MOSDEN separates the application-specific processing from the sensing, storing and sharing. MOSDEN is scalable and requires minimal development effort from the application developer.  We have implemented our framework on Android-based mobile platforms and evaluate its performance to validate the feasibility and efficiency of MOSDEN to operate collaboratively in mobile \osensing applications. Experimental outcomes and lessons learnt conclude the paper.

\end{abstract}


%
\IEEEpeerreviewmaketitle

\section{Introduction}
Today mobile phones have become a ubiquitous central computing and communication device in people's lives \cite{Lane-Survey}. The mobile device market is growing at a frantic pace and it wont be long before it outnumbers the human population. It is predicted that mobile phones combined with tablets will exceed the human population by 2017 \cite{web1}. Mobile phones more specifically smartphones are equipped with a rich set of on-board sensors, such as ambient light sensor, accelerometer, gyroscope, digital compass, GPS, microphone and camera. Moreover, current generation smartphones are equipped with technologies such as NFC, Bluetooth, WiFi that enable them to communicate and interact with external sensors available in the environment.

Smartphones have the potential to generate an unprecedented amount of data \cite{mobile-social-research} that can revolutionise many sectors of economy, including business, healthcare, social networks, environmental monitoring and transportation. According to Gartner\footnote{http://www.gartner.com/newsroom/id/2525515}, at present, smartphones dominate mobile phone sales with estimates indicating rapidly increasing smartphone shipments in the future. The data generated by an individual smartphone can be used to infer information about its user and to certain extent the environment around the user. By fusing data from a multitude of smartphones from a population of users, high level context information can be inferred. E.g., using an individual's smartphone, we can detect the current activity of the individual \cite{activity-recognition-mobile-sensing, thesis-wearable}. On the other hand, using data obtained from a population of individual's, we can detect the environmental context i.e. ambient light, noise in the environment \cite{mdm-paper}. In either form, the data generated by the smartphones are valuable and offers unique opportunities to develop novel and innovative applications. 

Most mobile sensing applications can be classified into \textit{personal} and \textit{community sensing} \cite{Lane-Survey, icdew-prem}. \textit{Personal sensing} applications focus on the individual. On the contrary, \textit{community sensing} also termed \textit{opportunistic/crowdsensing}\footnote{In this paper, we use the terms \textit{opportunistic sensing }, \textit{crowdsensing } and \textit{participatory sensing }synonymously.} takes advantage of a population of individuals to measure large-scale phenomenon that cannot be measured using single individual. In most cases, the population of individuals participating in \crowdsensing applications share a common goal. To date most efforts to develop \crowdsensing applications have focused on building monolithic mobile applications that are built for specific requirements \cite{pogo}. Further, the sensed data generated by the application are often available only within the closed population\cite{ganti}. However, to realise the greater vision of a collaborative mobile \crowdsensing application, we would need a common platform that facilitates easy development and deployment of collaborative crowd-sensed applications. 

The key challenge here is to develop a platform that is autonomous, scalable, interoperable and supports efficient sensor data collection, processing, storage and sharing. The autonomous ability of the system enables it to work independently when the device is off-line. Further, indiscriminately collecting all sensor data and transmitting it to a central server is expensive due to bandwidth and power consumption. We strongly believe that providing an easy to use, scalable platform to deploy collaborative mobile \crowdsensing applications will be significant for many new applications. To this end, we propose a collaborative mobile sensing framework namely Mobile Sensor Data Engine (MOSDEN). MOSDEN is capable of functioning on multitude of resource-constrained devices (e.g. Raspberry Pi\footnote{http://www.raspberrypi.org/}) including smartphones. MOSDEN is a scalable platform that enables collaborative processing of sensor data. The platform follows a component-based system paradigm allowing users to implement custom algorithms and models depending on application requirements. The key contributions of this paper are summarised as follows:

\begin{itemize}
\item We present the design and implementation of MOSDEN, a scalable, easy to use, interoperable platform that facilitates the development of collaborative mobile \crowdsensing applications
\item We demonstrate the ease of development and deployment using MOSDEN platform by demonstrating a collaborative mobile \crowdsensing application
\item We present experimental evaluation of MOSDEN's ability to respond to user queries under varying workloads to validate the scalability and performance of MOSDEN. 
\end{itemize} 

The rest of the paper is organised as follows. Section II discusses related work. Section III considers a motivation scenario. Section IV presents the proposed MOSDEN platform architecture. Section V discusses MOSDEN implementation and Section VI presents MOSDEN platform evaluations and results. Section VI concludes the paper with indicators to future work.

\section{Recent Work}

Mobile \crowdsensing popularly called community sensing \cite{ganti, oppur-mobile} is an autonomous collaborative sensing approach that requires minimal user involvement (e.g. continuous processing of noise level around user’s location). Numerous real and successful mobile crowd-sensing applications have emerged in recent times such as WAYZ\footnote{\label{note1}http://www.wayz.com/} for real-time traffic/navigation information and Wazer2\footnote{https://www.wazer2.co.il/}  for real-time, location-based citizen journalism, context-aware open-mobile miner (CAROMM)\cite{mdm-paper} among others. Mobile crowdsensing applications \cite{ZMP003, ZMP008} thrive on the data obtained from diverse sets of smart phones owned and operated by humans. Until recently mobile sensing application such as activity recognition (\textit{personal sensing)}, where people's activity (e.g. walking, talking, sitting) is classified and monitored, required specialised mobile devices \cite{activity-recognition-mobile-sensing, thesis-wearable}. This has significantly changed with advent of smartphones equipped with powerful computing, storage and on-board sensing capabilities. More recently, research efforts have focused on development of activity recognition, context-aware \cite{ZMP007} and data mining models on smartphones \cite{activty-recognition-gomes, minefleet, contextphone} that take advantage of smartphone's on-board sensing capabilities.

The efforts to build \crowdsensing application have focused on building monolithic mobile application frameworks that are built for specific purpose and requirements. Extending these frameworks to develop new applications is difficult, time-consuming and in some cases impossible. Crowd-sourcing data analytics system (CDAS) \cite{CDAS} is an example of a \crowdsensing framework. In CDAS, the participants are part of a distributed crowd-sensed system. The CDAS system enables deployment of various crowd-sensing applications that require human involvement for simple verification tasks delivering high accuracy. The system follows a two-stage approach. In the first stage, the given job is performed by a high-performance computer. The result of the job is then broken into subparts and sent to human workers for verification using Amazon Mechanical Turk (AMT). The results from human workers are combined to compute the final result. The CDAS system incorporates complex analytics that enables it to disseminate jobs, obtain results and compare results obtained from different workers to determine the correct one. Mobile edge capture and analysis middleware for social sensing applications (MECA) \cite{ye_meca:_2012} is another middleware for efficient data collection from mobile devices in a efficient, flexible and scalable manner. MECA provides a platform by which different applications can use data generated from diverse mobile data sources (sensors). The proposed MECA architecture has three layers comprising data layer (mobile data sources – mobile phones), edge layer (base stations that select and instruct a device or group of devices to collect data and process data), phenomena/application layer (the backend that determines the edge nodes to process application request). The mobile analytics performed on the data in CDAS and MECA are at the cloud/remote-server layer. 

\begin{figure*}[t!]
\vspace{-0.5cm}
\centering
\includegraphics[scale=0.4]{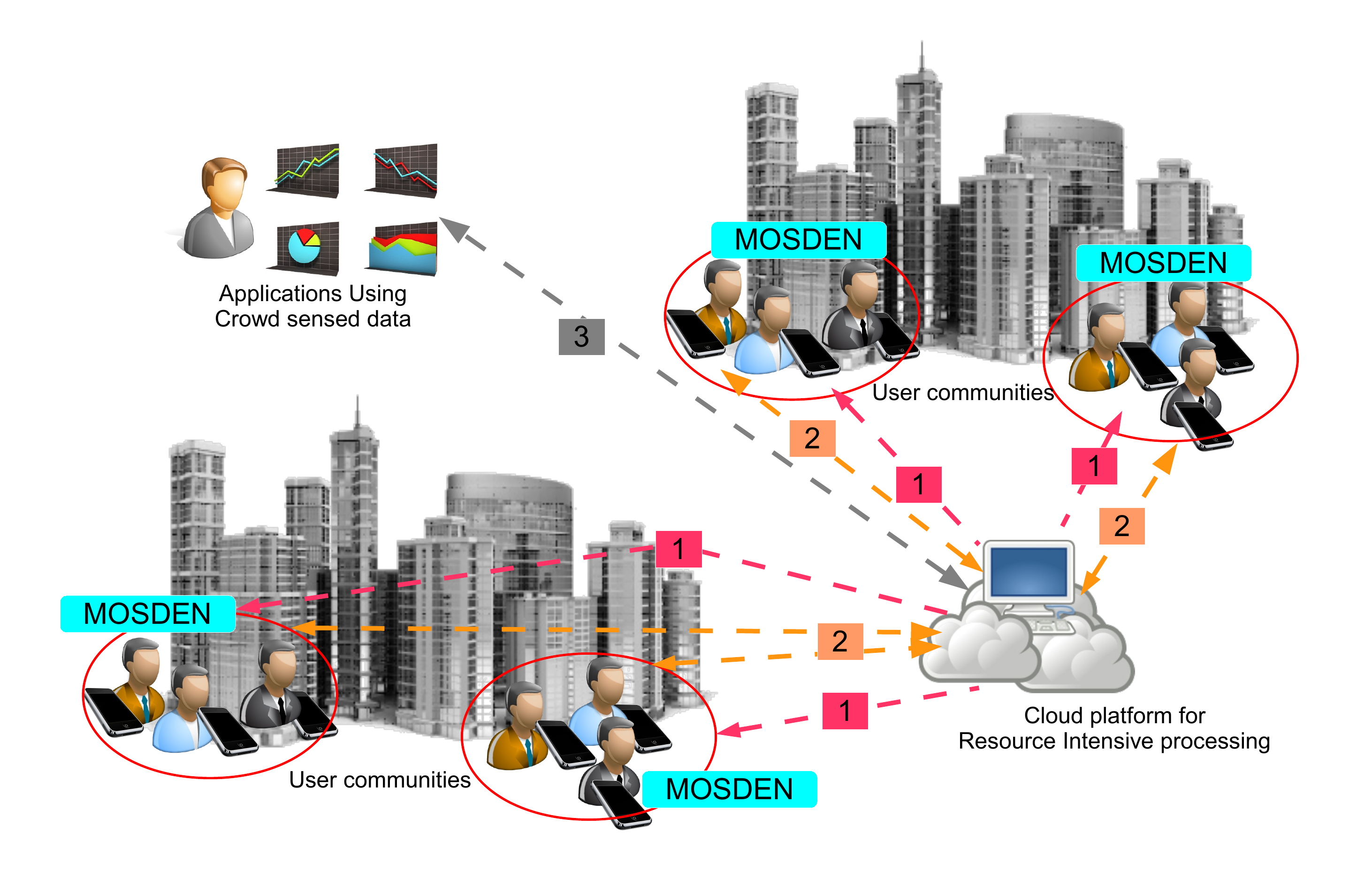}%
\vspace{-0.6cm}
\caption{Environmental Monitoring - Mobile Crowdsensing Scenario}
\label{scenario}
\vspace{-0.5cm}
\end{figure*}

The MetroSense \cite{metrosense} project at Dartmouth is an example of another \crowdsensing system. The project aims in developing classification techniques, privacy approaches and sensing paradigms for mobile phones. The CenceMe \cite{Miluzzo} project under the MetroSense umbrella is a personal sensing system that enable members of social networks to share their presence. 
The CenceMe application incorporates mobile analytics by capturing user activity (e.g., sitting, walking, meeting friends), disposition (e.g., happy, sad, doing OK), habits (e.g., at the gym, coffee shop today, at work) and surroundings (e.g., noisy, hot, bright, high ozone) to determine presence. The CenceMe system comprises two parts, the phone software and back-end software. The phone software is implemented on a Nokia N95 running Symbian operating system. The phone software is developed in Java Micro Edition (JME) which interfaces with Symbian C++ modules controlling the hardware. MineFleet \cite{minefleet} is a distributed vehicle performance data mining system designed for commercial fleets.  In MineFleet \cite{minefleet}, dedicated patented custom built hardware devices are used on fleet trucks to continuously process data generated by the truck. MineFleet system comprises an onboard data stream mining module that performs extensive processing of data using various statistical and data stream mining algorithms.  This data stored locally is transmitted to an external MineFleet Server for further processing when network connectivity is available. 

Mobile \crowdsensing is becoming a vital technique and has the potential to realise many applications that require large amounts of data from distributed communities in a collaborative fashion. The aforementioned \crowdsensing frameworks and applications are mostly hard wired allowing very little flexibility to develop new applications. Further, frameworks like MECA \cite{ye_meca:_2012} use the smartphone as a dumb data generator while all processing is offloaded to the server layer (Edge). This is good for certain types of applications but may not be suitable in scenarios where the smart phone may go off-line \cite{barriers-to-crowdsensing}. Moreover, \crowdsensing applications like Waze, MetroSense \cite{metrosense} and MineFleet \cite{minefleet} are built around specific data handling models (e.g. GPS for Waze, Microphone for MetroSense and Data mining algorithms for object monitoring). On contrast, the proposed MOSDEN platform has been developed with the design goal of ease of use, ease of development/deployment, scalability, easy access to both on-board and external sensors, support for on-board data analytics and collaboration and data sharing. The MOSDEN platform provides the application developer with implementation options i.e. choice of using processing on the smartphone and/or processing at the server. The MOSDEN platform promotes a distributed sensing infrastructure where each MOSDEN instance running on a smartphone is self-managed. 

\section{\label{section3} Motivating Scenario - Environmental Monitoring}

In this section we present a motivating futuristic scenario that explains the need for a scalable, collaborative, mobile sensing platform like MOSDEN. The scenario under consideration is an environmental monitoring scenario (e.g. noise pollution) in smart cities as depicted in Figure \ref{scenario}. In step (1), a remote-server (cloud-based) registers the interest for data within user communities. In the example depicted in Figure \ref{scenario}, the user communities are grouped based on location. In step (2), the processed data from the smartphones are sent to the remote-server (push/pull). In step (3), the crowd-sensing application obtains data from the remote-server for further processing and visualisation. The above scenario is a typical case for many \crowdsensing applications that require data from diverse user communities. The same approach can be used to deploy a \crowdsensing application that computes air pollution within the environment. To this, the smartphone will also have to rely on external sensors that are part of a smart city infrastructure to obtain air pollution data. Using a monolithic approach may results in developing a niche class of applications that may not be scalable for other scenarios which is a major obstacle. To achieve this goal, the \crowdsensing platform needs to support real-time data collection, processing and storage, support to implement specific algorithms/models, energy-efficient operation, autonomous functions i.e. ability to work with minimal user interaction and support offline modes. The proposed MOSDEN platform supports the above mentioned features natively.

\section {MOSDEN - MObile Sensor Data ENgine}
\label{section4}

We propose MOSDEN, a \crowdsensing platform built around the following design principles:
\begin{itemize}
\item Separation of data collection, processing and storage to application specific logic
\item A distributed collaborative \crowdsensing application deployment with relative ease
\item Support for autonomous functioning i.e. ability to self-manage as a part of the distributed architecture
\item A component-based system that supports access to internal and external sensor and implementation of domain specific models and algorithms
\end{itemize}

These design principles address the obstacles mentioned in Section \ref{section3}. The proposed MOSDEN platform overcomes the key barriers of developing and deploying scalable collaborative mobile \crowdsensing applications.

\subsection{Platform Architecture}

\begin{figure}[t]
\vspace{-0.3cm}	
 \centering
 \includegraphics[scale=0.38]{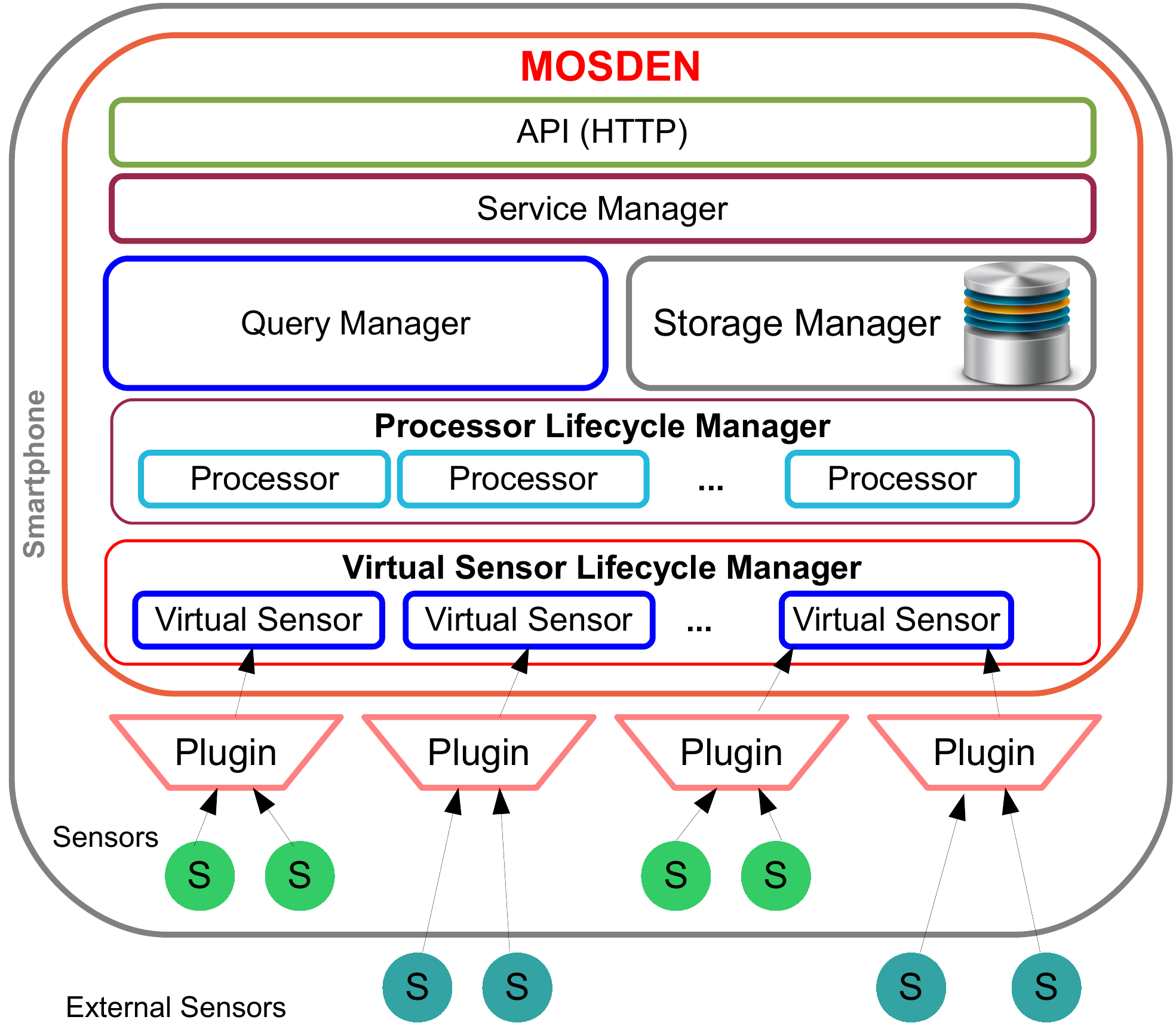}
 \caption{MOSDEN Platform Architecture}
 \label{sys-arch}	
\vspace{-0.8cm}	
\end{figure}

MOSDEN platform follows similar design principle of Global Sensor Network (GSN) architecture \cite{P227}. GSN is a sensor network middleware developed to run on high-powered computing devices (e.g. servers and cloud resources). GSN presents a unified middleware approach that facilitates acquisition, processing and storage of sensor data. It uses the concept of virtual sensors that abstracts the underlying data source (e.g. wireless sensor network). Since, GSN was not developed for resource constrained environment, we made significant enhancement to GSN when designing and implementing MOSDEN. MOSDEN follows a component-based architecture allowing extensibility without modifying the existing codebase. The architecture of the proposed MOSDEN platform is presented in Figure \ref{sys-arch} followed by description of each component.

\begin{figure}[h]
 \centering
 \includegraphics[scale=0.35]{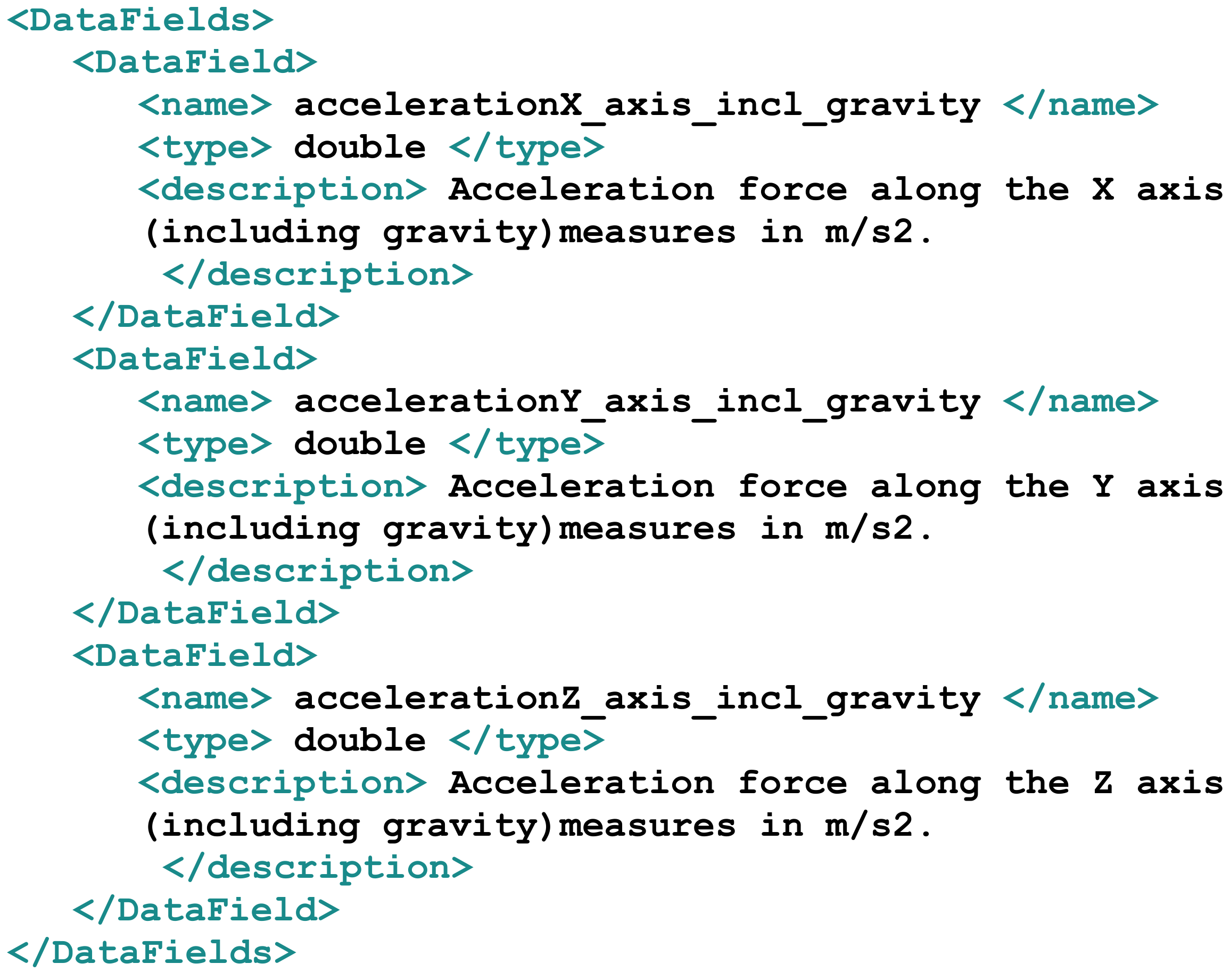}
 \caption{A Conceptual Description of MOSDEN Plugin}
 \label{plugin}	
\end{figure}

\begin{list} {\textit{Plugin:}}
\item  
In MOSDEN, we introduce the concept of Plugins. In GSN, a developer had to implement wrappers to accommodate new sensor data sources into the system. This required the system to be recompiled and redeployed. This approach is not very practical. The use of plugin overcomes this challenge. The Plugins are independent applications that communicates with MOSDEN. Plugin define how a sensor communicates with MOSDEN. We have developed a plugin descriptor that \crowdsensing application developer can use to implement plugins for the new sensor types. MOSDEN can dynamically discover new plugins at run-time. A conceptual description of the plugin is shown in XML format in Figure \ref{plugin}. 
\end{list}

\begin{list} {\textit{Virtual Sensor:}}
\item	The virtual sensor is an abstraction of the underlying data source from which data is obtained. This concept has been carried forward from GSN design. The virtual sensor lifecycle manager is responsible to manage the instantiation, updation and removal of virtual sensor resources.
\end{list}

\begin{list} {\textit{Processors:}}
\item	The processor classes are used to implement custom models and algorithms that processes the incoming data. For example, a Fast Fourier Transform (FFT) algorithm to compute the decibel level from microphone recordings.
\end{list}

\begin{list} {\textit{Storage Manager:}}
\item	The raw data acquired from the sensor is processed by the processing classes and stored locally. This is a key feature of MOSDEN as local storage supports off-line modes.
\end{list}

\begin{list} {\textit{Query Manager:}}
\item	The query manager is responsible to resolve and answer queries from external source. An external source can be another MOSDEN instance, a user or an application querying for data collected by the smartphone.
\end{list}
\begin{list} {\textit{Service Manager:}}
\item	The service manager is responsible to manage subscriptions to data from external sources. The service manager registers subscription request and depending on the mode of data delivery (push/pull) will deliver available data to the requested external source when possible. The service manager is specifically designed to manage the working on MOSDEN in resource constrained environments where frequent disconnection occurs.
\end{list}
\begin{list} {\textit{API Manager:}}
\item	The application programmable interfaces (APIs) provides a standard way to subscribe and access data to/from MOSDEN instances. The API's requests are received and processed over HTTP.
\end{list}

Each MOSDEN instance running on the mobile smartphones can run with minimal user interaction. It can register a data request from a remote-server (e.g. cloud-based). MOSDEN then works in the background processing the request by collecting, processing and storing the requested data locally. When the processed data is required by the application running at the remote-server, it can query the MOSDEN instance for the data (push/pull). MOSDEN realises a true distributed system architecture as it has the ability to function independent of the remote-server (support for off-line modes).

As depicted in the architecture, each individual MOSDEN instance is self contained and managed and is capable of working in mobile environments that encounter frequent disconnections. The use of APIs to communicate between instances encourages collaborative workload sharing and processing. The plugin based approach increases usability and promotes interoperability allowing MOSDEN to communicate with any sensors both internal and external. This remove the burden on \crowdsensing application developer. Further, the use of a component-based architecture enables system developers to implement domain specific algorithms with ease. Moreover, the MOSDEN platform enables the development of mobile \crowdsensing applications that can scale from an individual to a community. For example, the platform can be used to develop a personal fitness monitor application that works on an individual smartphone taking advantage of on-board sensing capabilities to noise pollution application that compute noise pollution by obtaining inputs from a community of users.

\section{Implementing a Crowdsensing Application using MOSDEN}

In Section \ref{section3} we presented an environmental monitoring scenario to determine the noise pollution level from data obtained from a community of user. Using the information obtained from the user communities, a \crowdsensing application running on a remote-server can further analyse and visualise the noise pollution level at a given location. Each user community in this scenario is grouped by location. 

\begin{figure}[t]
\vspace{-0.3cm}	
 \centering
 \includegraphics[scale=0.35]{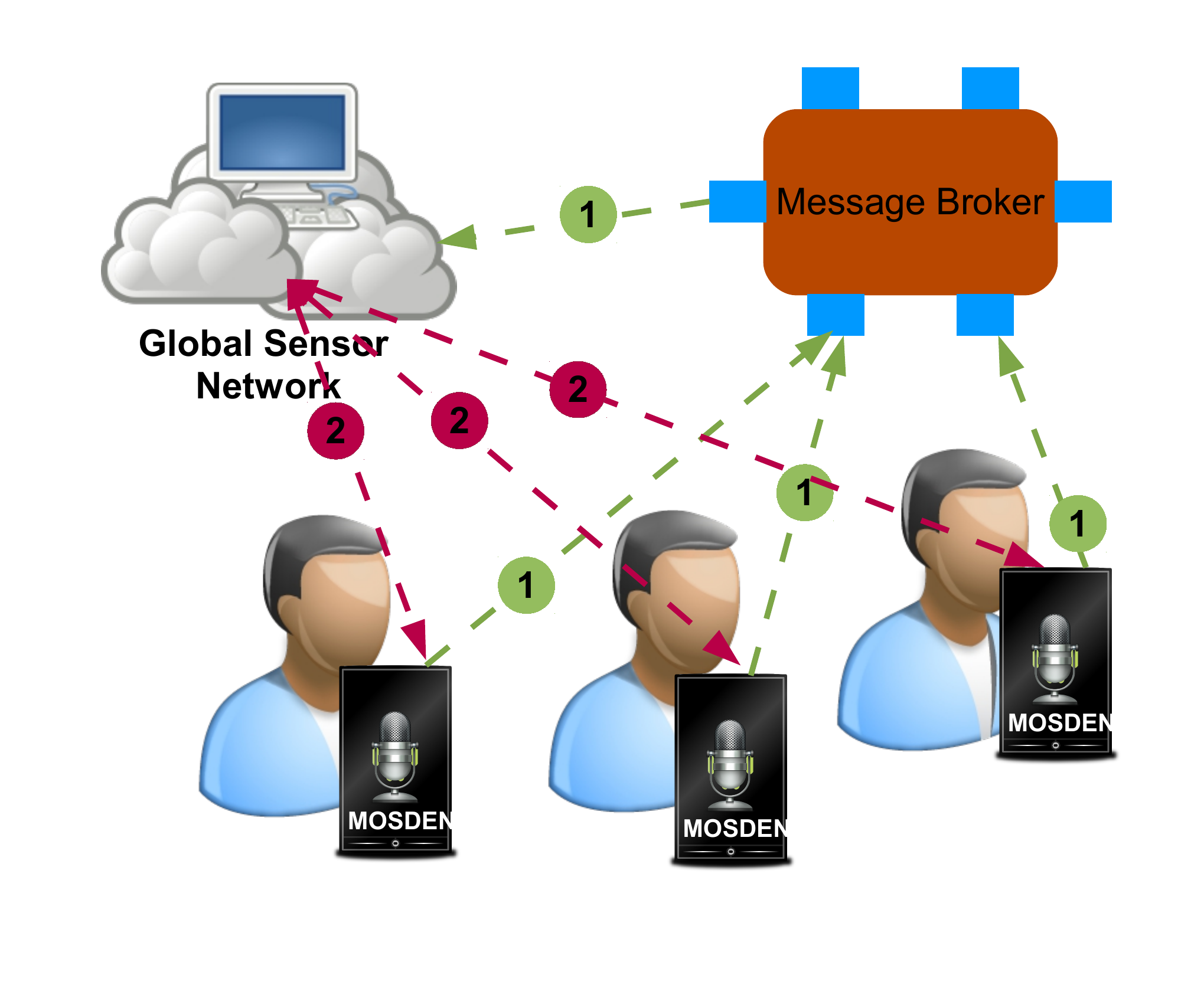}
\vspace{-0.6cm}	
 \caption{Implementation of Crowdsensing Application using MOSDEN}
 \label{Implementation of Crowdsensing Application using MOSDEN}	
\vspace{-0.8cm}	
\end{figure}

In this section we present a detailed description of the noise pollution \crowdsensing proof-of-concept application implementation using MOSDEN platform. Figure \ref{Implementation of Crowdsensing Application using MOSDEN} presents the overview of the noise pollution \crowdsensing application implemented on MOSDEN platform. In the scenario depicted in \ref{Implementation of Crowdsensing Application using MOSDEN}, in step (1) MOSDEN instances running on the smartphone registers with the cloud GSN instance. Once registration is complete in step (2) the cloud GSN instance registers its interest to receive noise data from MOSDEN. When data is available, MOSDEN streams the data to the cloud GSN. The streaming processes can be push or pull based depending on application requirement. In this specific example we implemented a pull-based approach.  

\begin{figure}
 \centering
 \begin{subfigure}[h]{0.4\textwidth}
  \centering
	 \includegraphics[width=\textwidth]{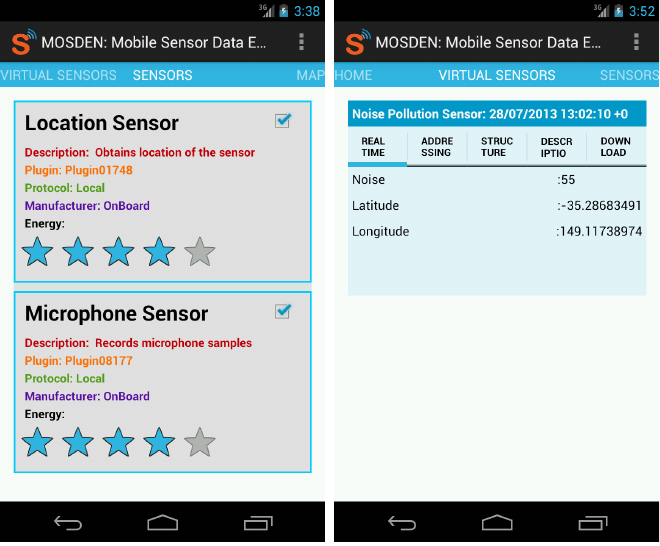}
	 \caption{MOSDEN User Interface}
	 \label{MOSDEN-screenshots1}	
 \end{subfigure}
 
 \begin{subfigure}[h]{0.4\textwidth}
  \centering
	 \includegraphics[width=\textwidth]{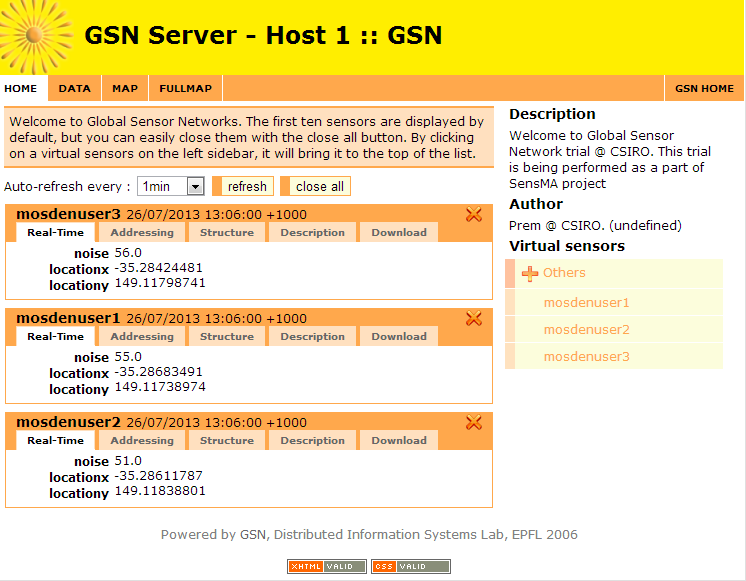}
	 \caption{GSN Sensor Registration Screenshot}
	 \label{GSN-screenshots1}	
 \end{subfigure}

 \begin{subfigure}[h]{0.4\textwidth}
  \centering
	 \includegraphics[width=\textwidth]{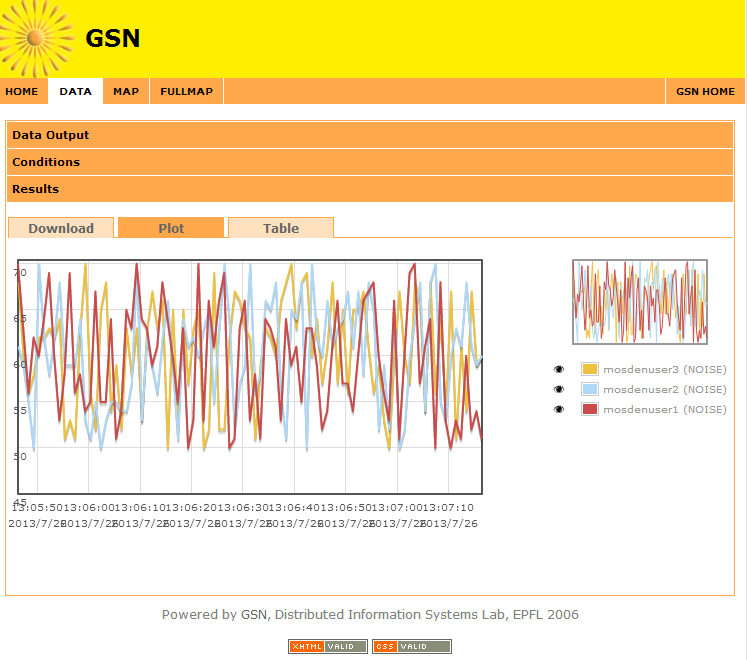}
	 \caption{GSN Noise Plot Screenshot}
	 \label{GSN-screenshots2}	
 \end{subfigure}
 \caption{Crowdsensing Application - Noise Pollution - Screenshots}
 \label{crowdsensing-app}
\vspace{-0.6cm}	
\end{figure}

The MOSDEN reference architecture has been implemented on the Android\footnote{http://www.android.com/} platform. We deployed the  noise pollution application developed on MOSDEN platform on a set of smartphones that represent user communities. To compute the noise decibel level, we implemented a modified version of the processing class from Audalyzer open source project\footnote{https://code.google.com/p/moonblink/}. The microphone sensor on the smartphones was used to obtain raw sound recordings. Code to interface with the sensor was already available as a part of the MOSDEN platform via plugins (we have developed plugins for on-board sensors). As MOSDEN is similar to GSN design, it is compatible with GSN. For our proof-of-concept implementation, we implemented GSN in the cloud that queries data from individual MOSDEN instances. A MOSDEN instance registers itself with the GSN in the cloud. As we stated earlier, the design of MOSDEN makes it easily extensible to suit any \crowdsensing application requirements. To validate this, we implemented the registration process via a message broker as depicted in Figure \ref{Implementation of Crowdsensing Application using MOSDEN}. Along with the registration, each MOSDEN instances also updates the cloud GSN instance with a list of available sensors. We note, MOSDEN API supports any form of registration. It is the responsibility of the \crowdsensing application developer to choose the most appropriate registration process. It is to be noted that the cloud GSN instance can be replaced by another smartphone running MOSDEN. In such a scenario, the MOSDEN requesting crowdsensed data performs further processing and visualisation. Screenshots of the MOSDEN implementation on Android smartphone (Figure \ref{MOSDEN-screenshots1}) and GSN in the cloud (Figure \ref{GSN-screenshots1}, \ref{GSN-screenshots2}) are illustrated in Figure \ref{crowdsensing-app}. We note, the default version of GSN with no enhancements was used to demonstrated the proof-of-concept implementation. Figure \ref{GSN-screenshots2} depicts the noise graph computed from 3 MOSDEN users. In this example, we have plotted the noise data individually.

\subsection{Benefits of MOSDEN Design}

The proposed MOSDEN model is architected to support scalable, efficient data sharing and collaboration between multiple application and users while reducing the burden on application developers and end users. The scalable architecture can easily be orchestrated for \crowdsensing application that range from an individual to a community of users. It facilitates easy sharing of data among large community of users which is a vital requirement for \crowdsensing applications.

By separating the data collection, storage and sharing from domain-specific application logic, our platform allows developers to focus on application development rather than understanding the complexities of the underlying mobile platform. In fact, our model hides the complexities involved in accessing, processing, storing and sharing the sensor data on mobile devices by providing standardised interfaces that makes the platform reusable and easy to develop new application. This we believe will significantly reduce the time to develop new innovative \crowdsensing applications. Since, MOSDEN is designed as a component-based architecture, it provides easy interfaces to implement application specific processing models and algorithms.

Further, our model works in the background with minimal user interaction reducing the burden on smartphone users. By providing support for processing and storage on the device, we also reduce frequent transmission to a centralised server as compared to current \crowdsensing frameworks. The potential reduction in data transmission has the following benefits: 1) helps save energy for users' mobile device; 2) reduces network load and avoids long-running data transmissions. 

To validate the performance of the proposed MOSDEN platform to support scalable, efficient data sharing and collaboration, in the next section, we evaluate the performance of MOSDEN to function under extreme loads when working collaboratively with other smartphones running MOSDEN instances. 

\section{Evaluation of MOSDEN Platform}
In this section, we present the details of experimentation test-beds and evaluation methodology. Further, we discuss the results and present the lessons learnt from  experimental evaluations.

\subsection{Experimentation Testbed}
\label{sec:E:Experimentation Testded}

For the evaluation of the proof of concept implementations, we used four mobile devices and a laptop. From here onwards we refer them as D1, D2, D3, D4, and D5 respectively. The technical specifications of the devices are as follows.

\begin{itemize}
\item \textbf{Device 1 (D1):} Google Nexus 4 mobile phone, Qualcomm Snapdragon S4 Pro CPU, 2 GB RAM, 16GB storage, Android 4.2.2 (Jelly Bean) 

\item \textbf{Device 2 (D2):} Google Nexus 7 tablet, NVIDIA Tegra 3 quad-core processor, 1 GB RAM, 16GB storage, Android 4.2.2 (Jelly Bean)

\item \textbf{Device 3 (D3):} Google Nexus 7 tablet, NVIDIA Tegra 3 quad-core processor, 1 GB RAM, 16GB storage, Android 4.2.2 (Jelly Bean)

\item \textbf{Device 4 (D4):} Acer Iconia Tab A501, Nvidia Tegra 2 T20 Dual-core 1 GHz Cortex-A9, 1 GB DDR2 RAM, Updated to  Android 4.2.2 (Jelly Bean), 

\item \textbf{Device 5 (D5):} ASUS Ultrabook Intel(R) Core i5-2557M 1.70GHz CPU and 4GB RAM (Windows 7 operating system)
\end{itemize} 

\begin{figure}[h]
 \centering
 \includegraphics[scale=0.35]{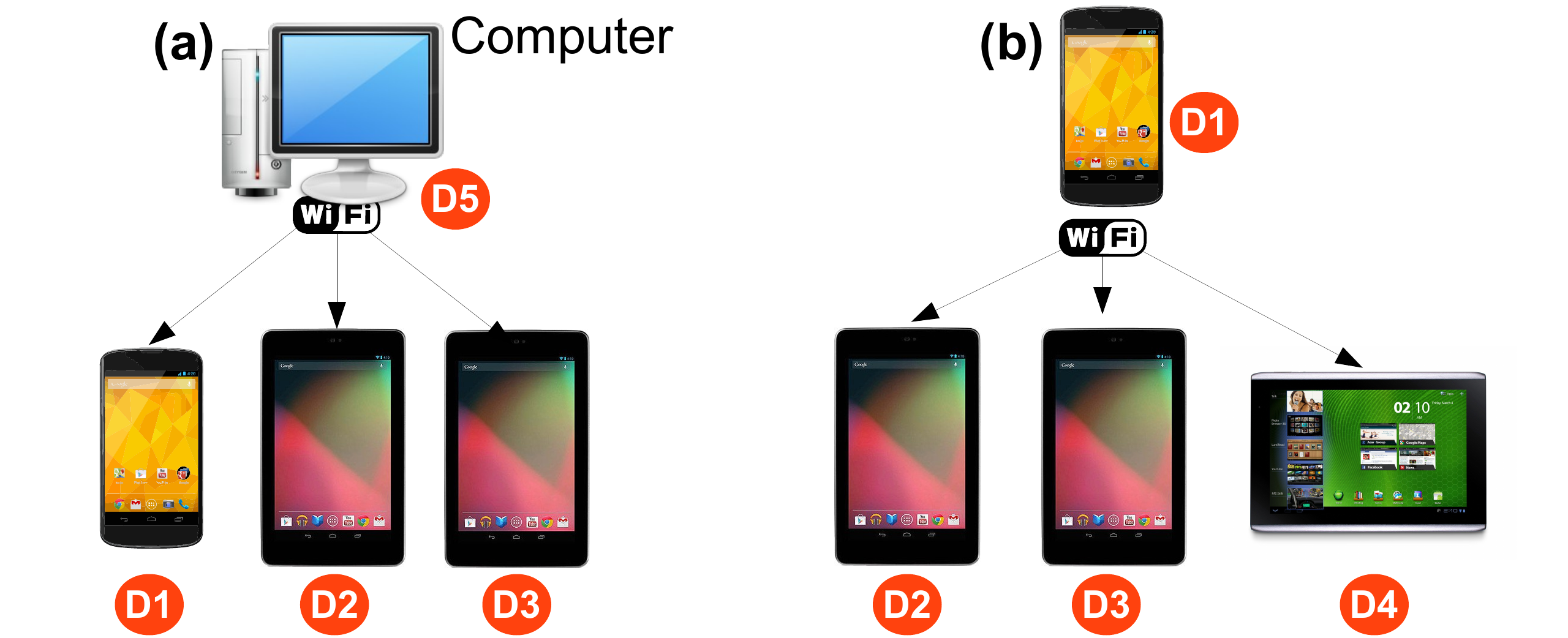}
\vspace{-0.22cm}	
 \caption{Experimental Testbed has been configured in two different ways: (a) Setup 1: Three mobile devices are connected to a laptop and 
(b)  Setup 2: three mobile devices are connected to another  mobile device.}
 \label{Figure:Setup}	
\vspace{-0.43cm}	
\end{figure}

For experimentation, we devised two setups as illustrated in Figure \ref{Figure:Setup} and evaluated the proposed framework in each setup independently. The mobile devices are configured to run our proposed framework, MOSDEN,  and the laptop computer is configured to run GSN engine \cite{P227}.

\subsection{Experimentation Strategy}
\label{sec:E:Experimentation_Strategy}

The overall objective of the experimental evaluations we conducted is to examine the performance of MOSDEN platform in collaborative environments. Two different collaborative setups are illustrated in Figure \ref{Figure:Setup}. In this section, we explain the objectives behind each experiment we conducted in detail. Next section discusses the results and lessons learnt in detail. Number of sensors used for sensing has been kept fixed throughout the experiments\footnote{All the sensors available on the given device has been used (e.g. In D1: accelerometer, microphone, light, orientation, proximity, gyroscope, magnetic, pressure).}. In all the evaluations, CPU usage (consumption) is measured in units of jiffies\footnote{In computing, a jiffy is the duration of one tick of the system timer interrupt. It is not an absolute time interval unit, since its duration depends on the clock interrupt frequency of the particular hardware platform}. Sampling rate for all evaluations is one second. 


A query in the form of a \textit{request} is sent from the server to MOSDEN client instances. Depending the number of sensors queried on MOSDEN instances, the number of requests increase. We use the term \textit{'MOSDEN client'} to refer to client devices where MOSDEN act as a client such as D1, D2 and D3 in setup 1 in Figure \ref{Figure:Setup}(a) and D2, D3 and D4 in setup 2 in Figure \ref{Figure:Setup}(b)). We use the term  \textit{'MOSDEN server'} to refer to server device where MOSDEN act as a server such as D1 in setup 2 in Figure \ref{Figure:Setup}(b)).

We configured the experimental test-bed as illustrated in Figure \ref{Figure:Setup}(a) - Setup 1. In Figure \ref{Experiment1}, \ref{Experiment2}, and \ref{Experiment3}, we compare the performance of \textit{restful streaming} and \textit{push-based streaming} methods in term of CPU usage and memory usage by both client and server devices which runs MOSDEN and GSN. Restful streaming is designed to have a persistent connection between the client and the server. On the other hand, the push-based approach makes a new connection every time to transmit data. Both these techniques can be used to perform communication between two (or more) distributed GSN or MOSDEN instances (i.e. GSN $\leftrightarrow$ GSN, MOSDEN $\leftrightarrow$ MOSDEN,  GSN $\leftrightarrow$ MOSDEN). The two approaches have their own strengths and weakness. The former is good for clients running MOSDEN that have a reliable data connection. The latter is useful for clients that need to work in offline modes. The MOSDEN platform supports both the operations and the application developer has the choice to choose the best approach suited to application requirements.

Figure \ref{Experiment1} illustrates the difference between CPU usage in MOSDEN when number of requests increase. Figure \ref{Experiment2} illustrates the variation of memory consumption of MOSDEN when number of requests increase. Figure \ref{Experiment3} illustrates how memory consumption of GSN changes in the server when number of queries it handles increase. 

In Figure \ref{Experiment4}, we examine  how storage requirements vary when number of sensors handled by the MOSDEN client increases. For this experiment, we used Setup 1 in \ref{Figure:Setup}. All the sensors onboard the client mobile device (i.e. accelerometer, microphone, light, orientation, proximity, gyroscope, magnetic, pressure) are used as sensor sources. Sampling rate for sensors are configured as one second. The D1 (Setup 1) has been configured to receive data request from the server in an one second interval. The experiment was conducted for three hours. The exact storage requirements depend on multiple factors such as number of active sensors sending data, number of data items generated by the sensor\footnote{E.g. accelerometer generates 3 data items i.e. x, y, and z while temperature sensor generate one data item}, sampling rate, and history size \cite{P022}. We used external sensor to increase the number of sensors connected to MOSDEN during the experiment in order to examine the behaviour of MOSDEN from a storage requirement perceptive.

For the next set of experiments, we configured the test-bed as illustrated in Figure \ref{Figure:Setup}(b)-Setup 2. In Figure \ref{Experiment5} and \ref{Experiment6}, we compare the performance of restful streaming and push-based streaming techniques in terms of CPU usage and memory usage by the server mobile device (D1) which runs MOSDEN.  Figure \ref{Experiment5} illustrates the difference between CPU usage in MOSDEN when number of requests increase. Figure \ref{Experiment6} illustrates the variation of memory consumption of MOSDEN when number of requests increase. 

Figure \ref{Experiment7} shows how round trip time\footnote{The round-trip time is the time taken for the server to request a data item from a given virtual sensor on a client. The total time is computed as the interval elapsed between server request and client response.} is impacted when the number of requests handled by GSN (D5 in Figure \ref{Figure:Setup}(a)) and MOSDEN (D1 in Figure \ref{Figure:Setup}(b)) increase. Both restful streaming and push-based streaming techniques are evaluated separately. Figure \ref{Experiment8}, compares the amount of time (average) it takes to process a single request\footnote{Time taken to process a single request is the time interval elapsed between two subsequent requests made by the server to any client irrespective of the virtual sensor}. This is different from round trip time presented in  Figure \ref{Experiment7}. Time it takes to process a single request is calculated as denoted in Equation \ref{equation1}.

\begin{equation}
 = \frac{\textrm{Duration of the Experiment}}{\textrm{Total number of Round Trips Completed}}
\label{equation1}
\end{equation}

In Figure \ref{Experiment9}, we presents results of our experiment (Figure \ref{Figure:Setup}-Setup 2) that examine how each request was processed. We compared the performance using both restful streaming and push-based streaming. In this experiment, we configured MOSDEN server to make 30 requests from each of the three distributed client MOSDEN instances. We conducted the experiment for a fixed interval of time. Later, we calculated, using the Equation \ref{equation2}, the number of round-trips completed by each request and plotted them as a percentage. We denote the total number of round-trip requests completed for a virtual sensors $S$ as $S_{i}$ where $i$ is the virtual sensor identifier. The x-axis in Figure \ref{Experiment9} represents $i$.

\begin{equation}
 = \left ( \frac{\textrm{Number of Round trips Completed by } S_{i}}{\textrm{Total number of Round Trips Completed}  \sum_{i=1}^{n} S_{i}} \right ) \times 100
\label{equation2}
\end{equation}

In Figure \ref{Experiment10}, we visually illustrate how delay occurs in processing the 90 requests (in Figure \ref{Experiment9}, we only show 7 requests due to space limitation). Each request is shown in a different colour. Different requests have different round-trip times depending on how processing capabilities and priorities of both server and client devices.

\subsection{Results and Discussion}
\label{sec:E:Results_and_Discussion}

\begin{figure}[h]
 \centering
 \includegraphics[scale=0.45]{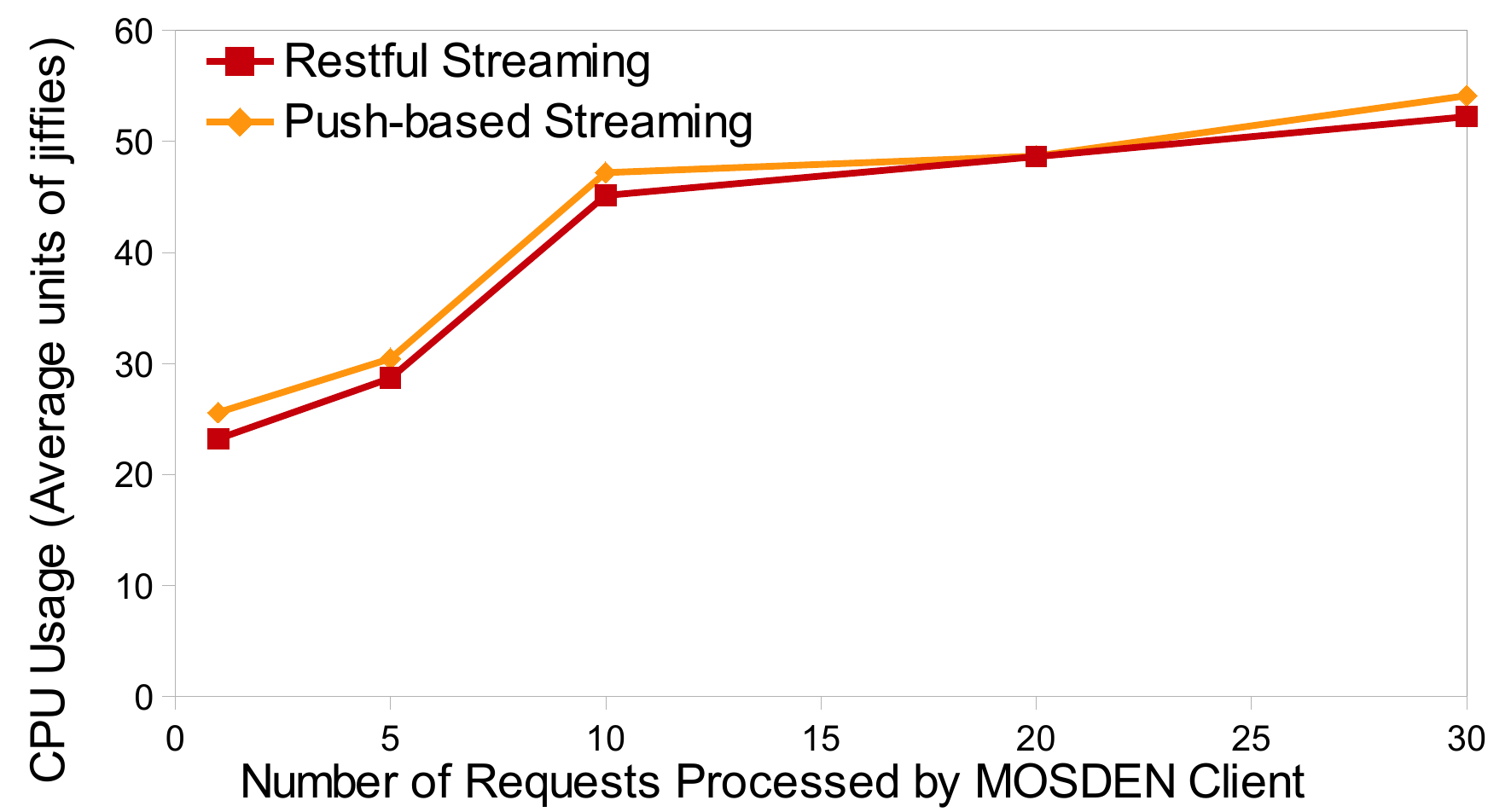}
 \caption{Comparison of CPU Usage by MOSDEN Client}
 \label{Experiment1}	
\vspace{-0.43cm}	
\end{figure}

\begin{figure}[h]
 \centering
 \includegraphics[scale=0.45]{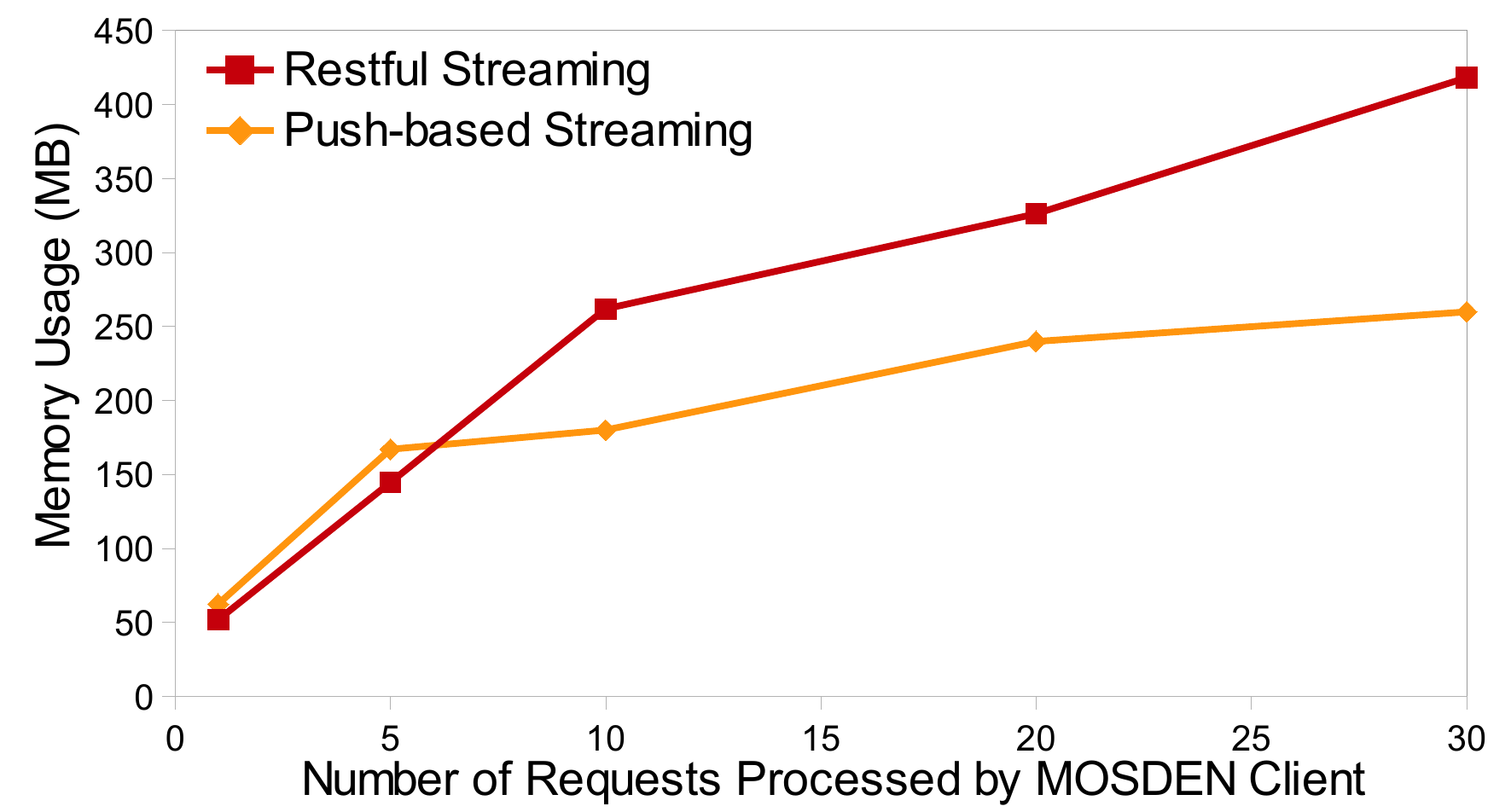}
 \caption{Comparison of Memory Usage by MOSDEN Client}
 \label{Experiment2}	
\vspace{-0.43cm}	
\end{figure}

In this section, we provide a detailed analysis and discussion of the experimental outcomes. According to Figure \ref{Experiment1}, it is evident that restful streaming is slightly better than push-based streaming in CPU consumption perceptive. This slight different can be due to above explained reasons. On contrast, restful streaming consumes more memory than push-based streaming as depicted in Figure \ref{Experiment2}. One reason could be the overheads to maintain a persistent network connections.

\begin{figure}[h]
 \centering
 \includegraphics[scale=0.45]{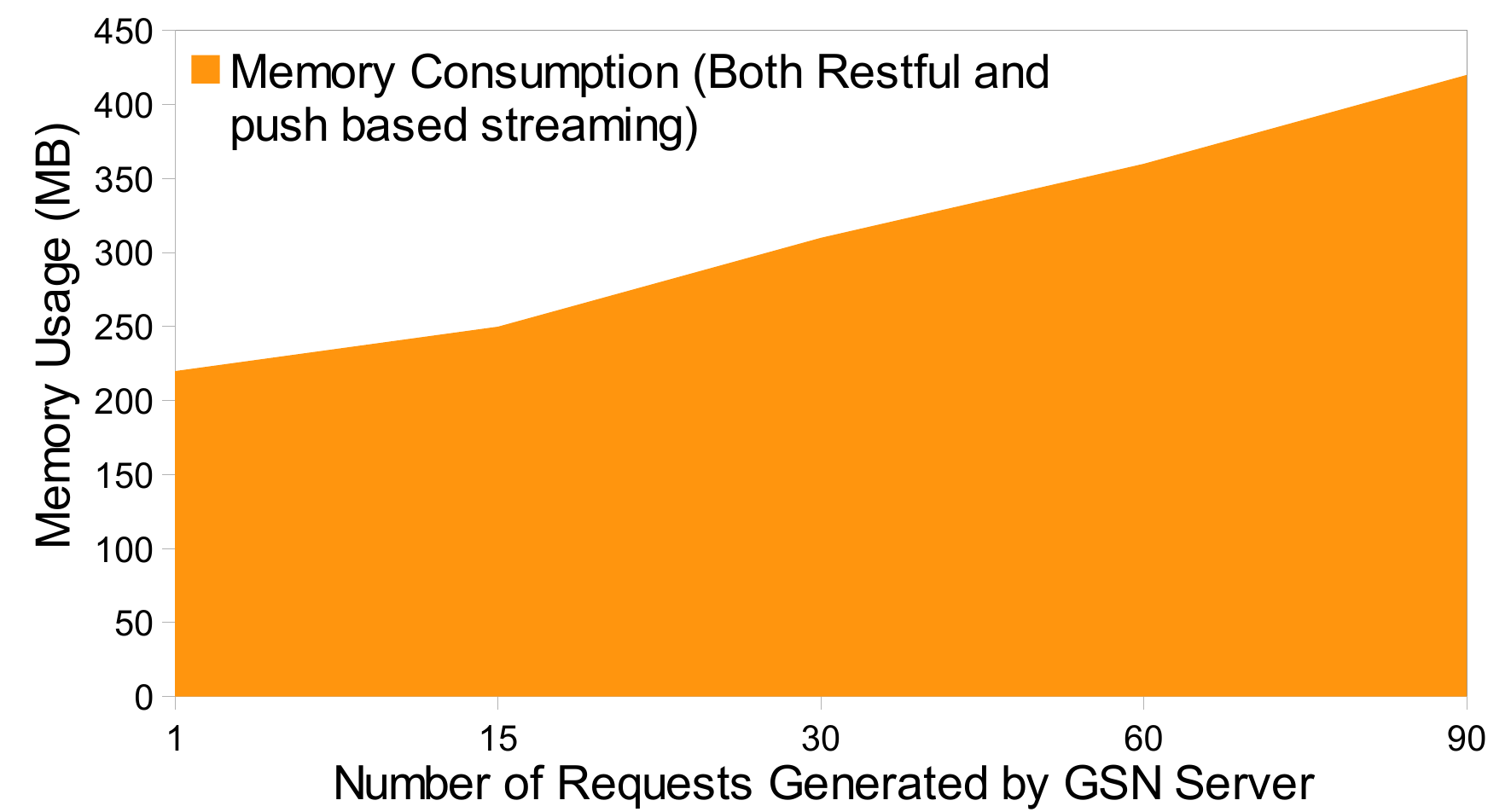}
 \caption{Comparison of Memory Usage by GSN Engine}
 \label{Experiment3}	
\vspace{-0.43cm}	
\end{figure}

\begin{figure}[h]
 \centering
 \includegraphics[scale=0.45]{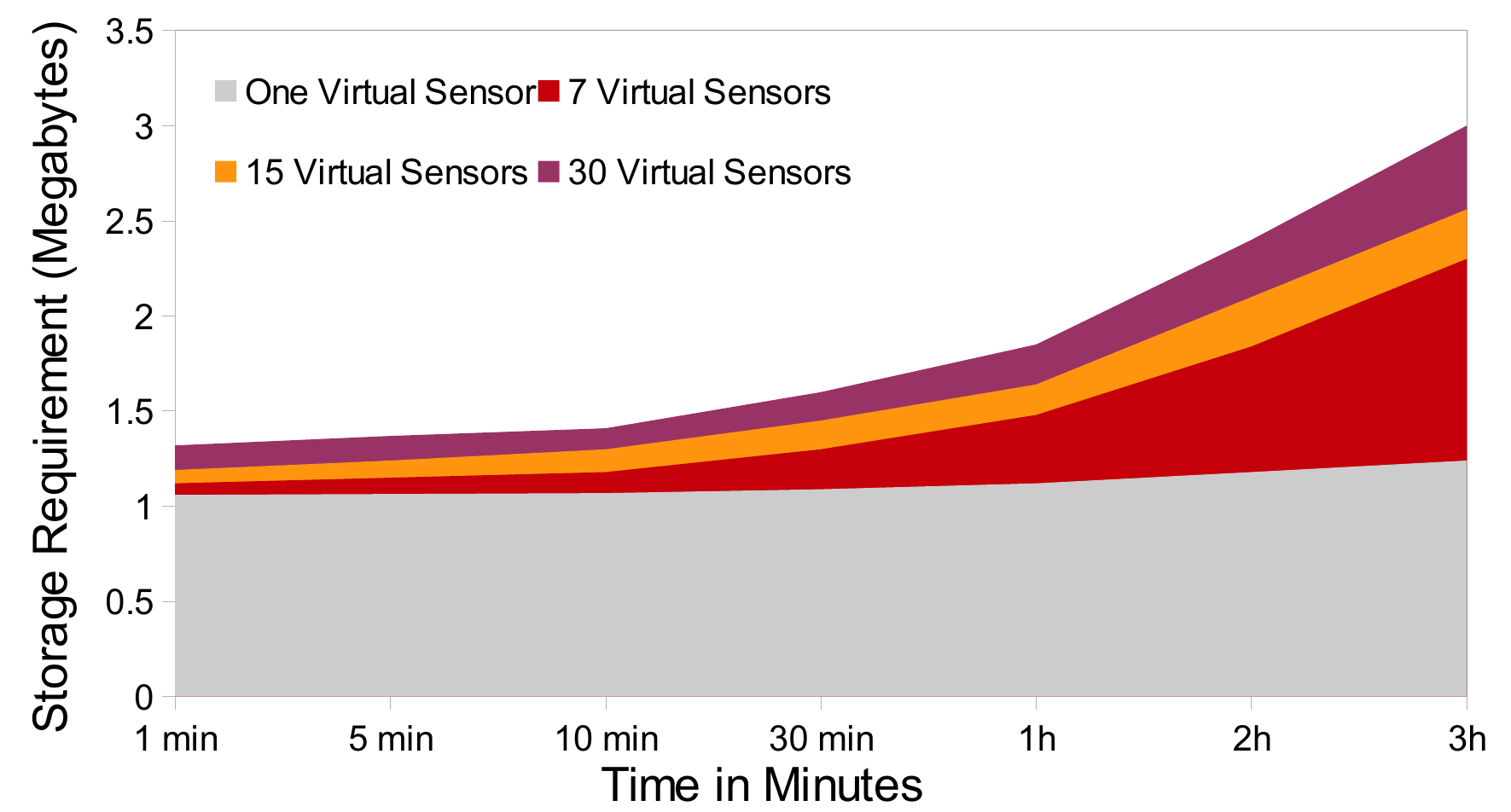}
 \caption{Storage Requirement of MOSDEN}
 \label{Experiment4}	
\vspace{-0.43cm}	
\end{figure}

It can also be noted that the memory consumption of GSN engine running on the server as depicted in Figure \ref{Experiment3} also increases with load but not as significant as the mobile device. This observation is straightforward attributed to the difference in computing capacity of the two nodes (mobile device and laptop). Based on the experience in MOSDEN client-side, it is fair to predict that, we will be able to see a different if we increase the number of requests to be processed towards tens of thousands. According to the outcome shown in Figure \ref{Experiment4}, storage  requirements are linear. It is to be noted that to stress test MOSDEN client instances, we used external sensors, on-board sensors and additional data source generators to simulate 30 virtual sensors. This further demonstrates the scalability of MOSDEN. In both GSN and MOSDEN, storage can be easily controlled by changing the history-size. History-size defines how much data record needs to be stored at a given time. Large history sizes can be used for summarising purposes or archival purposes. However, the amount of storage in easily predictable due to history size, because MOSDEN always deletes old items in order to accommodate new data items. Specially, for real time reasoning history can be set to one. Considering all the above factor, it is fair to conclude that modern mobile devices have the storage capacity to store sensor data collected over long period of time.

\begin{figure}[h]
 \centering
 \includegraphics[scale=0.45]{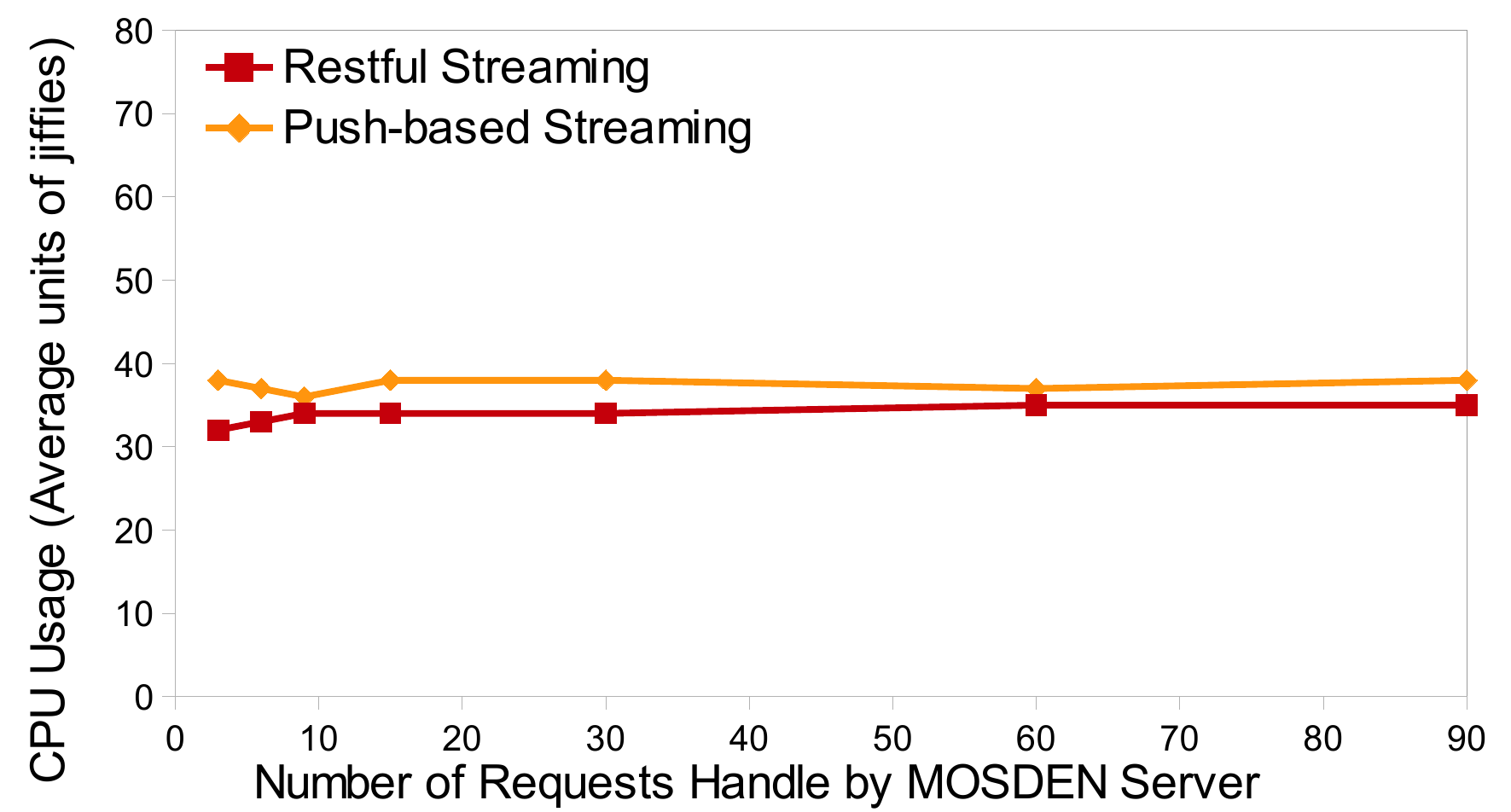}
 \caption{Comparison of CPU Usage by MOSDEN Server}
 \label{Experiment5}	
\vspace{-0.43cm}	
\end{figure}

\begin{figure}[h]
 \centering
 \includegraphics[scale=0.45]{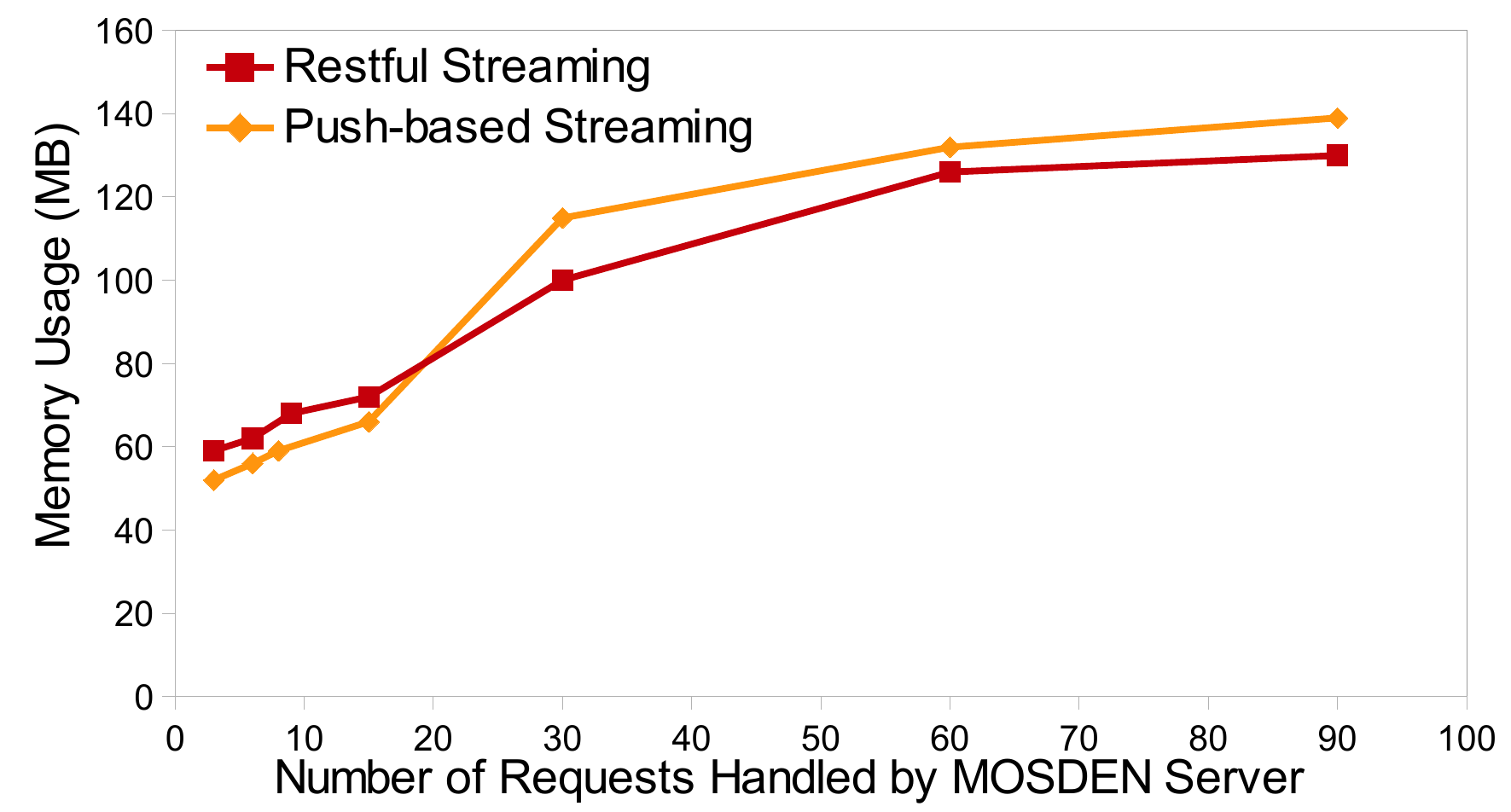}
 \caption{Comparison of Memory Usage by MOSDEN Server}
 \label{Experiment6}	
\vspace{-0.43cm}	
\end{figure}

\begin{figure}[h]
 \centering
 \includegraphics[scale=0.45]{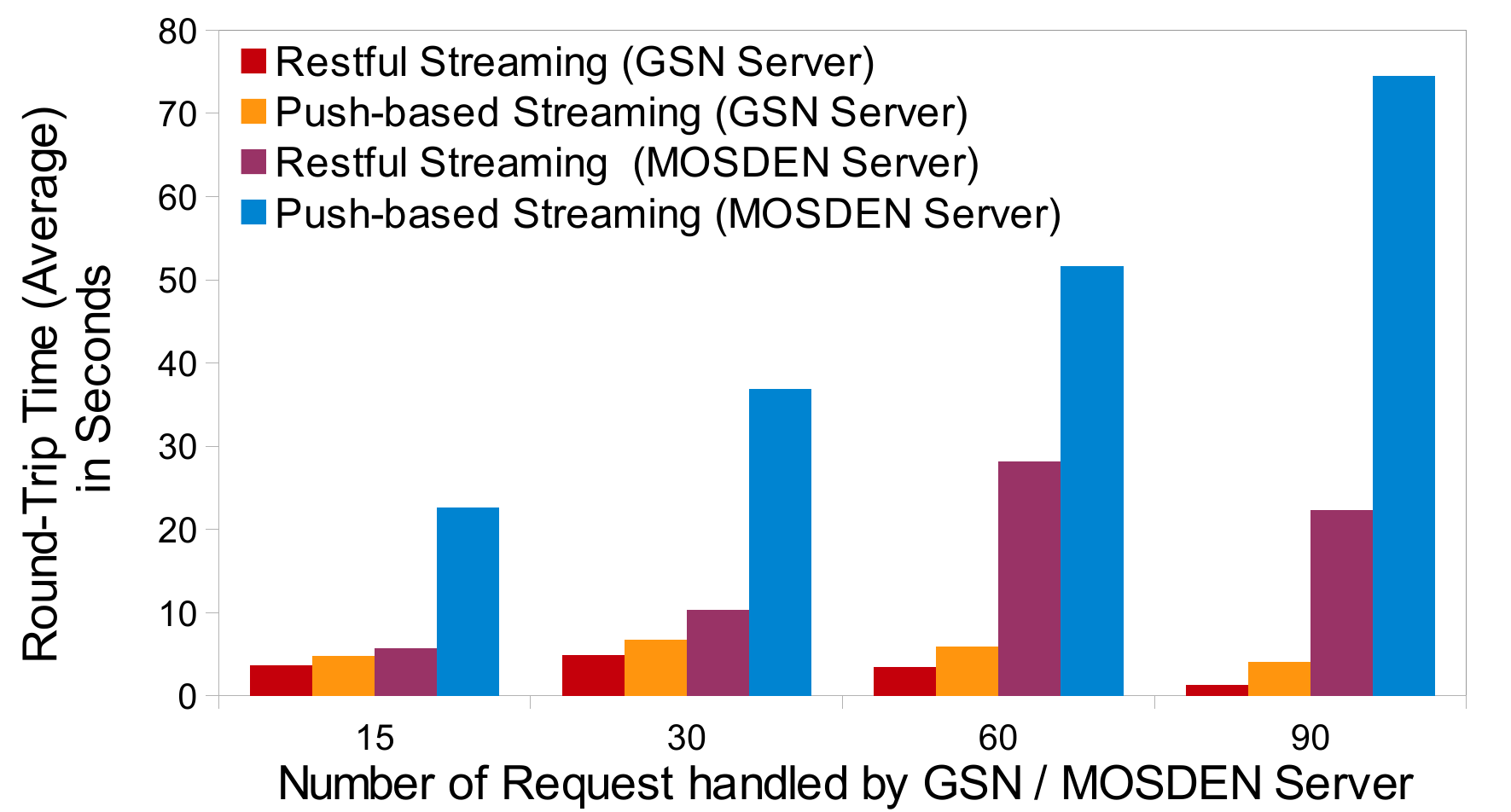}
 \caption{Comparison of Round-trip Times}
 \label{Experiment7}	
\vspace{-0.43cm}	
\end{figure}

According to Figure \ref{Experiment5} and Figure \ref{Experiment6} Push based streaming is slightly better that restful streaming. Further, it is important to note that both techniques maintain the same amount of  CPU consumption over time despite the increase in requests in handles. Additionally, MOSDEN server consumes significantly less amount of memory in comparison to MOSDEN client. One reason is that MOSDEN client performs sensing activities in addition to sending data to the server. In contrast, MOSDEN server  performs data requesting task only (from clients). As we mentioned earlier, when number of requests handled by MOSDEN increase (give that no other tasks are performed), restful streaming technique performs better in term of both CPU consumption and memory consumption.

\begin{figure}[t]
 \centering
 \includegraphics[scale=0.45]{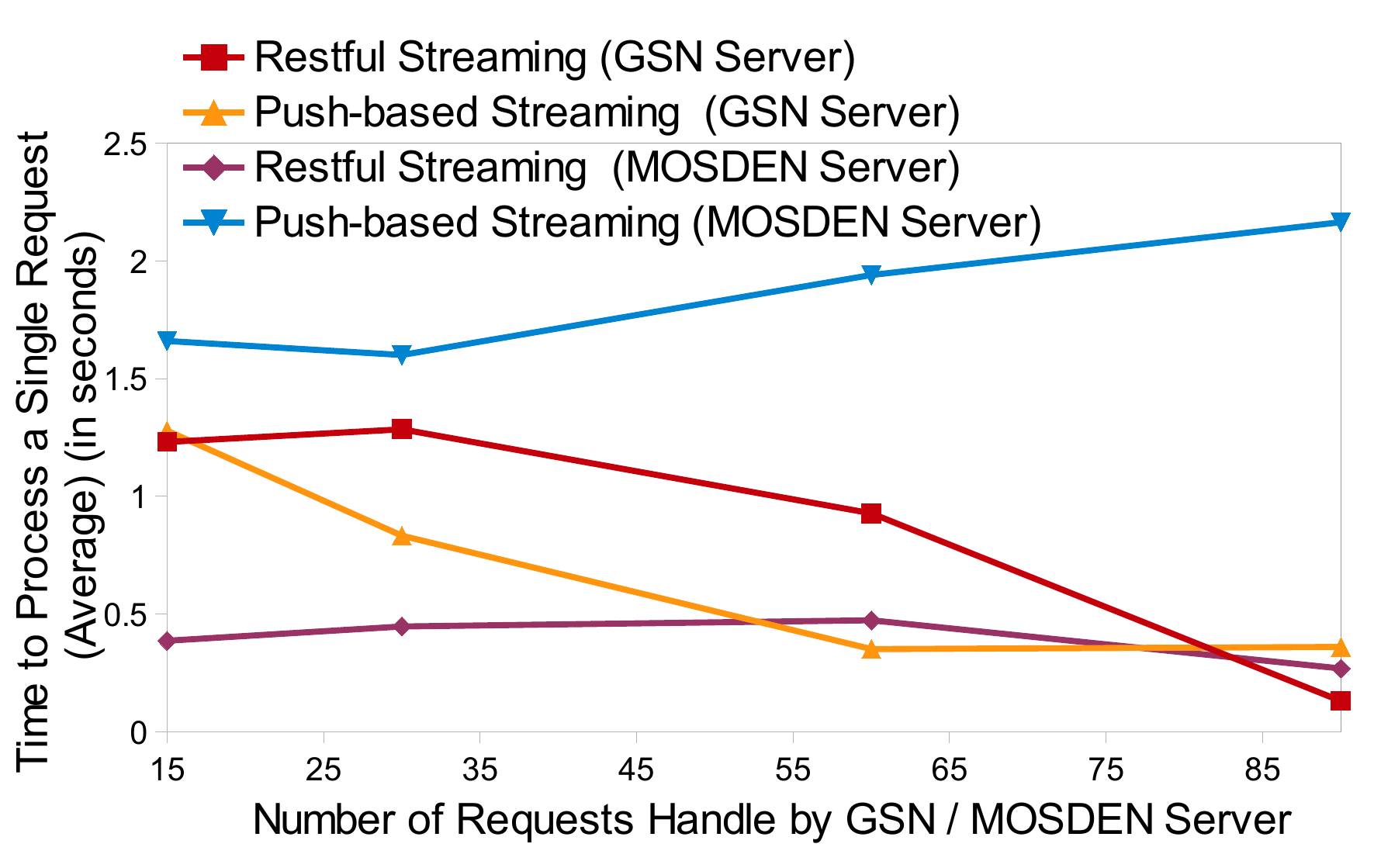}
 \caption{Comparison of Data Retrieval and Processing Ability}
 \label{Experiment8}	
\vspace{-0.8cm}	
\end{figure}

\begin{figure*}[b]
 \centering
 \includegraphics[scale=0.45]{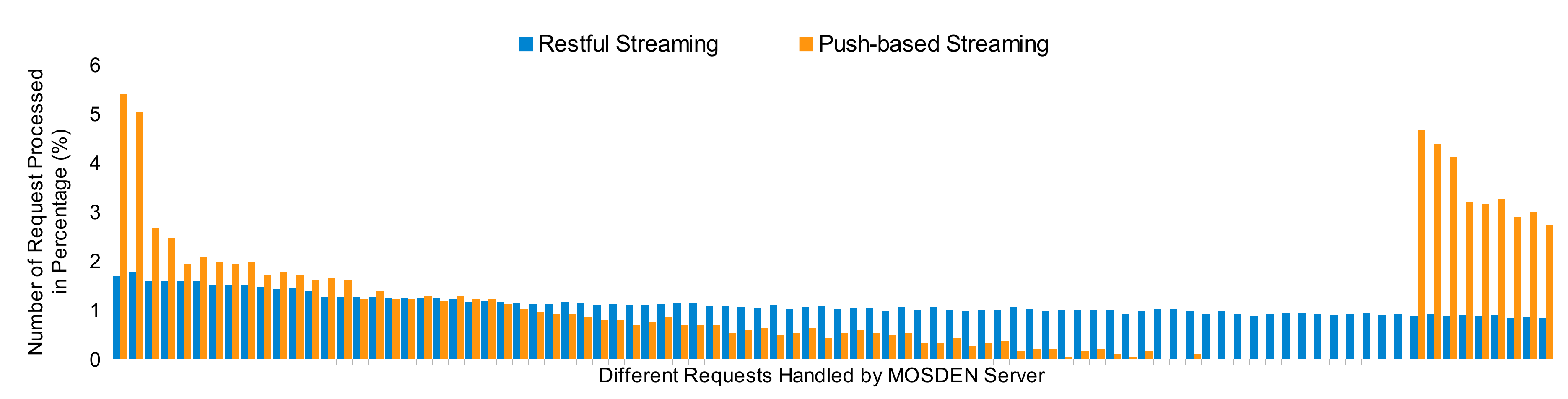}
\vspace{-0.22cm}	
 \caption{Comparison of Requests Processing Variation}
 \label{Experiment9}	
\vspace{-0.7cm}	
\end{figure*}

\begin{figure*}[b]
 \centering
 \includegraphics[scale=0.45]{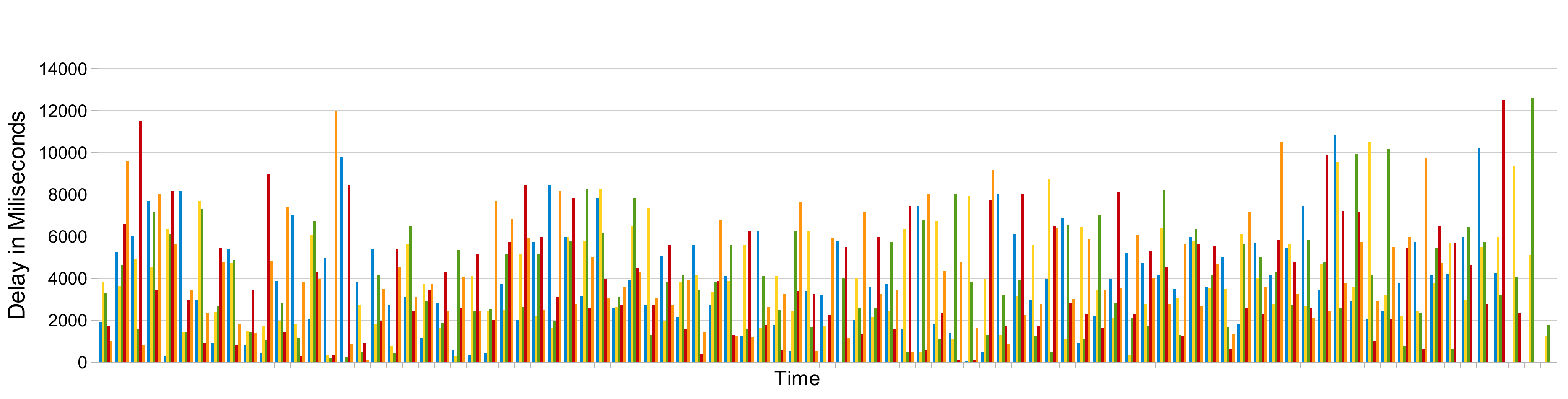}
\vspace{-0.22cm}	
 \caption{Variation of round-trip time (delay / latency) over a period of time where seven requests are being processed}
 \label{Experiment10}	
\vspace{-0.43cm}	
\end{figure*}

According to Figure \ref{Experiment7}, it is clearly evident that resource constrained device such as mobile phones take more time to perform computations. As a result delay time is comparatively high when the server node is a mobile device in contrast to a computer-based server node. Further it has been observed that (also we predicted in earlier section), push-based technique has much larger delay time due to additional overheads involved in connection setup and teardown. For laptop-based server instances, the reason for having much less round trip time when handling 90 requests is due to the availability of more computational resources. However, when resource constrained devices play the role of a server node, they do not have additional CPU or memory to allocate in comparison to a laptop-based server. As a result round trip time increases for mobile device-based server nodes. Figure \ref{Experiment8} also shows the impact of increased overheads when using a push-based streaming technique. 


It is important to note that, even though, the average round trip time is  higher as observed in Figure \ref{Experiment7}(e.g. 20 seconds when handling 90 requests) when restful steaming techniques is used, the amount of time taken to make subsequent requests by the server is mush less (e.g. less than a second when handling 90 requests) as observed in Figure \ref{Experiment8}. This outcomes is explained in Figure \ref{Experiment9}. As some virtual sensor requests complete more round trips (as explained earlier) compared to others, the delay introduced to process request for other virtual sensors have a direct impact on average round trip time. 

According to Figure \ref{Experiment9}, restful streaming technique allows each request to have fair amount of computational resources but push-based streaming does not. The main reason is attributed to the fact that restful streaming maintains a persistent connection between the client and server. When devices use push-based streaming, more computational resource needs to be allocated to handle the connection setup and teardown. Specially, when the number of requests that need to be handled increase significantly, it places a significant overhead on round-trip times for the push-based streaming approach as shown in Figure \ref{Experiment9}. Due to restricted resources, under extremely high loads, in push-based streaming, there is a fair possibility that some requests made by virtual sensors (in MOSDEN server) may not get executed at all. In Figure \ref{Experiment10}, we plotted round trip times taken by 7 different requests over a period of time. This clearly shows the significance of  the variation stated above. Some requests (in some point of time) take only 6 milliseconds whereas some other requests (in some point of time) take 12 seconds to complete a round trip.

Overall MOSDEN performs extremely well in both server and client roles in collaborative environments. MOSDEN (as a server) was able to handle 90 requests (i.e. 180 sub requests) where each request has a sampling rate of one second. This resulted in a MOSDEN clients processing 1800 data points every 1 minute and a MOSDEN server (running on a mobile device) processing 5400 data points every 1 minute from distributed clients. It is to be noted, that for evaluation purposes and to validate the efficiency and scalability of MOSDEN, we conducted experiments on MOSDEN server and client under extreme loads. Such processing is intensive and rare in real-world application. However, our experiments showed that MOSDEN can withstand such intensive loads proving to be a scalable platform for deploying large-scale \crowdsensing applications. If MOSDEN is configured to collect data from 10 different sensors and handle 30 requests (typical of real-world situations), it can perform real-time sensing with delay of 0.4 - 1.5 seconds. When the server node is a computer (D5 as explained in Section \ref{sec:E:Experimentation Testded}) both restful streaming and push-based streaming work extremely well without visible significant differences. However, when the server node is a mobile device, which runs MOSDEN, restful streaming performs approximately 6 times better than push-based technique. 

%

\section{Conclusion and Future Work}
A mobile \crowdsensing application development framework must scale from an individual user to user communities (100 -1000 users). In this paper, we proposed MOSDEN, a collaborative mobile \crowdsensing platform to develop and deploy \osensing applications. MOSDEN differs from existing \crowdsensing platforms by separating the sensing, collection and storage from application specific processing. This unique feature of MOSDEN renders it an easy-to-use, reusable framework for developing novel \osensing applications. We proposed the architecture of the MOSDEN framework. We then demonstrated its ease of use and minimal development effort by presenting a proof-of-concept noise pollution application developed on the MOSDEN platform. We validated MOSDEN's performance and scalability when working in distributed collaborative environments by extensive evaluations under extreme loads resolving and answering queries from external sources (MOSDEN instances and GSN in the cloud). Overall MOSDEN performs extremely well under extreme loads in collaborative environments validating its suitability to develop large-scale \osensing applications. Our next step is to deploy and evaluate MOSDEN in a real-world application.


\section*{Acknowledgement}

Part of this work has been carried out in the scope of the ICT OpenIoT Project which is co-funded by the European Commission under seventh framework program, contract number FP7-ICT-2011-7-287305-OpenIoT. The authors acknowledge help and support from CSIRO Sensors and Sensor Networks Transformational Capability Platform (SSN TCP).

\bibliography{Bibliography}
\bibliographystyle{IEEEtran}

%
%
%

\end{document}